\documentclass[12pt]{iopart}
\usepackage[latin1]{inputenc}
\usepackage[T1]{fontenc}
\usepackage{iopams}
\usepackage{graphicx}
\usepackage{cite}
\usepackage{siunitx}
\usepackage{tikz}
\usepackage{subfigure}

\usepackage{color}

\begin{document}
\sisetup{
	retain-explicit-plus = true
}
\title{On the synthesis of Quantum Hall Array Resistance Standards}
\author{Massimo Ortolano\textsuperscript{1,3}, Marco Abrate\textsuperscript{2} and Luca Callegaro\textsuperscript{3},
		\\[\medskipamount]  % Authors lines
		     {\small \textsuperscript{1} Dipartimento di Elettronica e Telecomunicazioni, Politecnico di Torino } \\
        {\small Corso Duca degli Abruzzi 24, 10129 Torino, Italy} \\
        {\small \textsuperscript{2} Dipartimento di Ingegneria Meccanica e Aerospaziale, Politecnico di Torino} \\
        {\small Corso Duca degli Abruzzi 24, 10129 Torino, Italy} \\
        {\small \textsuperscript{3}INRIM - Istituto Nazionale di Ricerca Metrologica} \\
        {\small Strada delle Cacce, 91 - 10135 Torino, Italy} \\
}

\begin{abstract}
Quantum Hall effect (QHE) is the basis of modern resistance metrology. In Quantum Hall Array Resistance Standards (QHARS), several individual QHE elements, each one having the same QHE resistance (typically half of the von Klitzing constant), are arranged in networks that realize resistance values close to decadic values (such as \SI{1}{\kilo\ohm} or \SI{100}{\kilo\ohm}), of direct interest for dissemination. The same decadic value can be approximated with different grades of precision, and even for the same approximation several networks of QHE elements can be conceived. The paper investigates the design of QHARS networks by giving methods to find a proper approximation of the resistance of interest, and to design the corresponding network with a small number of elements. Results for several decadic cases are given.
\end{abstract}

\maketitle
\section{Introduction}
Quantum Hall Array Resistance Standards (QHARS) \cite{Piquemal:1999, Poirier:2002, Bounouh:2003, Poirier:2004, Hein:2004, Oe:2008, Oe:2010, Oe:2011, Konemann:2011, Woszczyna:2012, Oe:2013} are integrated circuits in which several quantum Hall elements are interconnected to compose a resistive network. The so-called \emph{multiple-series} and \emph{multiple-parallel} connections (see \cite{Ortolano:2012} and references therein) allow the realization of four-terminal resistances with negligible effect from contact and wiring resistances. 

National metrology institutes maintain and disseminate the unit of electrical resistance through artifact resistance standards having decadic nominal values, such as \SI{100}{\ohm}, \SI{1}{\kilo\ohm}, \SI{10}{\kilo\ohm}. Resistance metrology would strongly benefit from the development of reliable QHARS having resistance values close to decadic values, because the calibration of artifact standards could thus be performed by 1:1 ratio bridges, which do not require ratio calibration, or even by substitution.

QHARS require dedicated foundries and the development of novel realization methods, in particular for what concerns wiring and insulation. Although the present integration level allows the realization of QHARS with hundreds of elements, a basic goal of QHARS design is to maximize simplicity, keeping the number of the elements to a minimum.

Recent papers have shown that the same approximation of a decade resistance value $R$ can be realized with networks having a greatly different number of elements; for example, in the realization of a $\SI{10}{\kilo\ohm}$ QHARS, the National Metrology Institute of Japan (NMIJ) evolved from a 266-element network~\cite{Oe:2008} to a 16-element one~\cite{Oe:2013}.

The aim of the present paper is to give advice on the synthesis of QHARS networks which approximate a given resistance value within the accuracy required in resistance metrology and employing a minimal number of elements. 

\section{The problem}
A QHARS is a circuit composed of interconnected elements having the same resistance $R_\textup{H} = R_\textup{K}/i$, where $R_\textup{K}$ is the von Klitzing constant and $i$ is the \emph{plateau index} (typically, $i=2$). The total resistance $R$ of a QHARS is a fraction of the element resistance, $R = (p/q) R_\textup{H}$, where $p$ and $q$ are positive integers which depend only on the network topology. 

The main goal in the design of a QHARS is to obtain a device having a resistance value close to a target value $R_0$. In the following, for ease of notation, we consider the normalized dimensionless quantities $\rho_0 = R_0/R_\textup{H}$ (a generic real number) and $\rho =  R/R_\textup{H} = p/q$ (a rational number).   

The problem of QHARS network synthesis can be divided into the following steps:
\begin{enumerate}
\item Find a rational approximation $\rho = p/q$ of $\rho_0$ such that the magnitude of the relative error $\delta = (\rho-\rho_0)/\rho_0$ is less than a specified limit $\delta_\textup{max}$, that is $|\delta| < \delta_\textup{max}$.
\item Synthesize an optimal network with normalized resistance $\rho$ with a minimal number $n$ of elements.
\end{enumerate}

\section{Rational approximations}
\begin{figure}
\centering
\includegraphics[width=\textwidth,clip=]{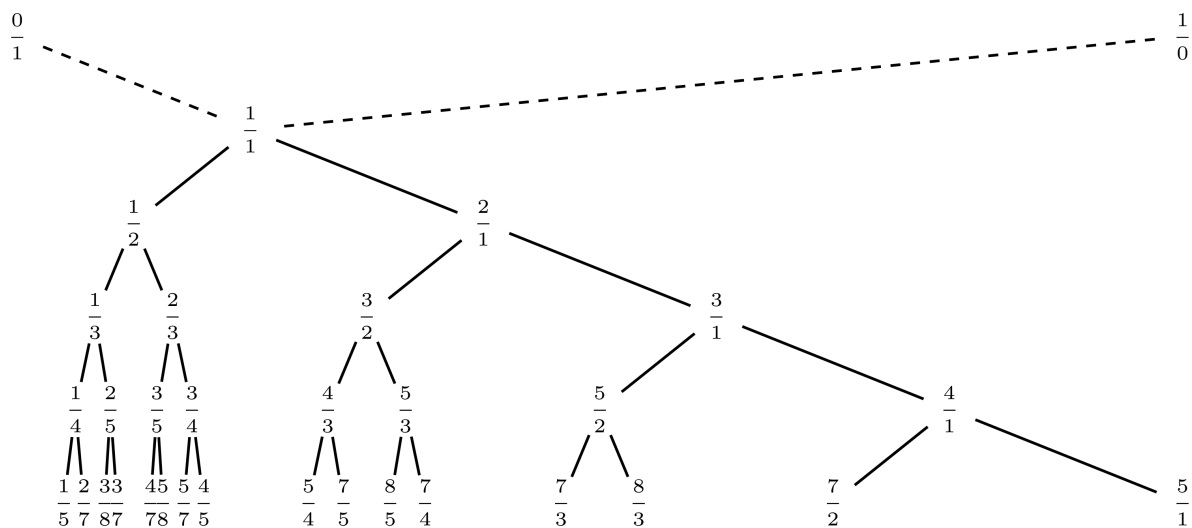}
\caption{Top levels of the Stern-Brocot tree (for fractions different from $1/0$ the horizontal position is to scale).}
\label{fig:sb_tree}
\end{figure}

In this section we consider the problem of obtaining a rational approximation for a given positive real number. Historically, this problem was of particular interest to clockmakers, who had to design accurate gear trains, and led the French clockmaker Achille Brocot to conceive the useful construction which is discussed below\footnote{Before the advent of direct digital frequency synthesizers, time and frequency metrologists ---the modern clockmakers--- confronted with this problem too (see e.g.~\cite{Kroupa:1974}).}. The so-called Stern-Brocot tree~\cite{Stern:1858,Brocot:1862,Graham:1994} is a binary tree whose nodes are in one-to-one correspondence with the set of positive rational numbers. In particular (figure~\ref{fig:sb_tree}), each node is the mediant fraction $(a+c)/(b+d)$ of the left and right nearest ancestors $a/b$ and $c/d$ (the fractions that can be reached by following the branches upward); the root node is the fraction $1/1$, generated from the fictitious ancestors $0/1$ and $1/0$. 

\begin{figure}
\centering
\includegraphics[width=\textwidth,clip=]{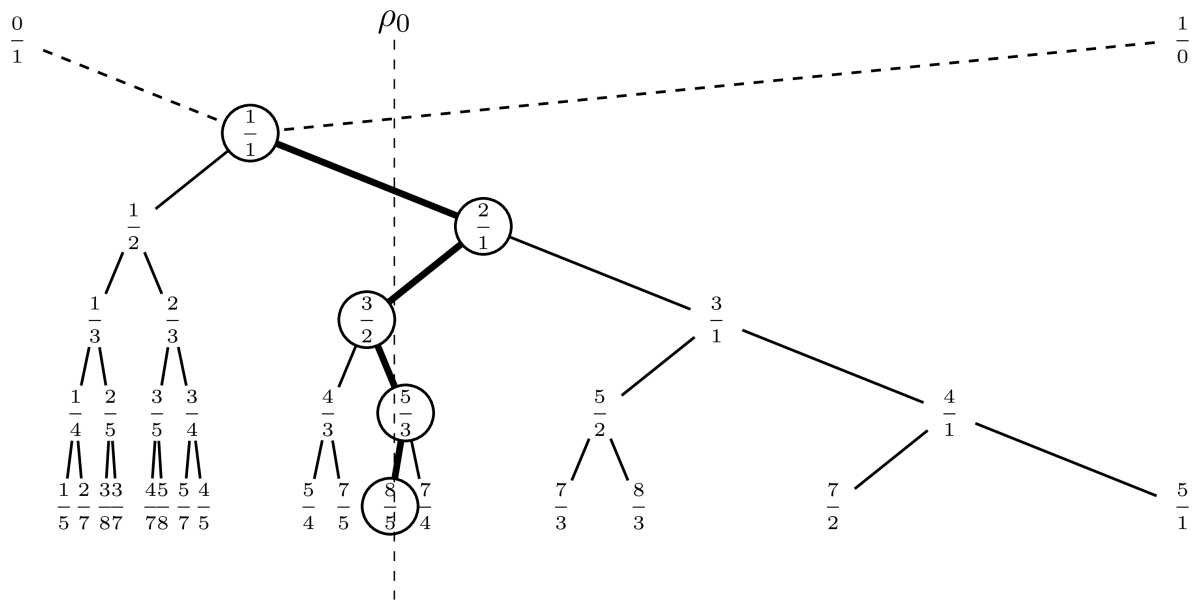}
\caption{Walking down the Stern-Brocot tree to approximate a real number $\rho_0$ (here the Stern-Brocot sequence is $\frac{1}{1},\frac{2}{1},\frac{3}{2},\frac{5}{3},\frac{8}{5},\ldots$).}
\label{fig:sb_treewalk}
\end{figure}

The importance of Stern-Brocot tree in the approximation theory rises up by considering the sequences of rationals, namely the Stern-Brocot sequences, obtained by walking down the nodes of the tree.  Given a real number $\rho_0$, the corresponding Stern-Brocot sequence $\rho_1,\rho_2,\rho_3,\ldots$ can be generated by walking down the tree according to the following rules (see figure~\ref{fig:sb_treewalk}): start from the tree root $\rho_1 = 1/1$; then, for each $k = 1,2,3,\ldots$, if $\rho_k < \rho_0$, set $\rho_{k+1}$ equal to the left child of $\rho_k$; otherwise, if $\rho_k > \rho_0$, set $\rho_{k+1}$ equal to the right child; if, for some $k$, $\rho_k = \rho_0$, stop (if $\rho_0$ is rational, the sequence is finite). The Stern-Brocot sequence $\rho_1,\rho_2,\rho_3,\ldots$ obtained in this way converges to $\rho_0$; moreover, the terms of the sequence are rational approximation of $\rho_0$ that are optimal in the following sense: if a certain rational approximation of $\rho_0$ does not belong to the sequence, than there is a term of the sequence having smaller numerator or denominator which is a better approximation~\cite{Graham:1994}. 

A known method to approximate real numbers is that of continued fraction expansion. A continued fraction is a representation of a real number $\rho_0$ through a sequence of integers as follows:
\begin{equation}
\rho_0 = \alpha_0+\frac{1}{\alpha_1+\displaystyle{\frac{1}{\alpha_2+\displaystyle{\frac{1}{\alpha_3+\cdots}}}}}\ ,
\end{equation}
where the integers $\alpha_0,\alpha_1,\ldots$ are given by the recurrence relations~\cite{Olds:1963} 
\begin{eqnarray}
\xi_0=\rho_0\,, \\
\alpha_k = [\xi_k]\,,\\	
\xi_{k+1}=\frac{1}{\xi_k-\alpha_k}\,, \quad\text{if $\xi_k$ is not an integer},
\end{eqnarray} 
where $k = 0,1,2,\ldots$, and $[\xi_k]$ denotes the integer part of $\xi_k$. A continued fraction can be expressed in a compact way using the notation $[\alpha_0;\alpha_1,\alpha_2,\alpha_3,\ldots]$. 

The finite continued fractions
$[\alpha_0;\ldots,\alpha_m]$, for $m = 0,1,2,\ldots$, are rational approximations of $\rho_0$. Given a real number $\rho_0$, all the rational approximations obtained by continued fraction expansion are included in the Stern-Brocot sequence associated to the same $\rho_0$: therefore, approximations obtained by continued fraction expansion need not to be considered separately.

\section{QHARS network synthesis}
\label{sec:qhars_synthesis}
This section considers the problem of synthesizing a planar resistive network from a given rational approximation $\rho = p/q$ of the target resistance $\rho_0$. In the following, we consider $p$ and $q$ to be coprimes. Furthermore, since we are dealing with normalized quantities, all elements are of unit value. The systematic generation of such networks was investigated in~\cite{Amengual:2000,Khan:2012,OEIS:A153588}.

\begin{figure}[htbp]
 	\centering
	\includegraphics{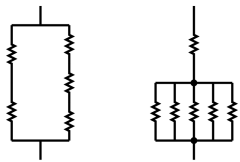}
 	\caption{Two different networks with $\rho = 6/5$. The network on the left has $n=5$ elements, while that on the right has $n=6$ elements.}
 	\label{fig:net_6x5}
\end{figure}

For a certain $\rho$, several different equivalent networks having a different number $n$ of elements can be synthesized. A simple example is shown in figure~\ref{fig:net_6x5}, where $\rho = 6/5$ is obtained with two different networks. 

\begin{figure}
\centering
\begin{minipage}[b]{0.25\textwidth}
\centering
\includegraphics{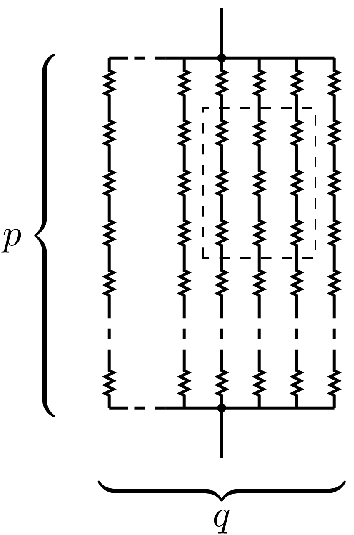} 
\end{minipage}
\hspace{2ex}
\begin{minipage}[b]{0.25\textwidth}
\centering
\includegraphics{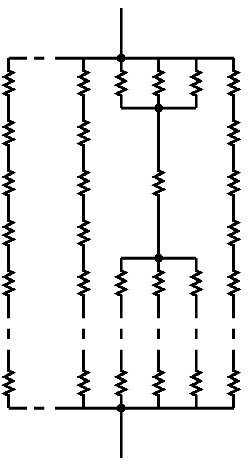} 
\end{minipage}
\\
\begin{minipage}[t]{0.25\textwidth}
\centering
(a) 
\end{minipage}
\hspace{2ex}
\begin{minipage}[t]{0.25\textwidth}
\centering
(b)
\end{minipage}	
\caption{(a) The rectangular $p \times q$ network. (b) A $3\times 3$ square of elements has been substituted with a single element.}
\label{fig:net_pxq}
\end{figure}

A basic network which realizes $\rho=p/q$ in a trivial way is the rectangular network of figure~\ref{fig:net_pxq}, which is composed of $n=p q$ elements arranged in a rectangular grid of height $p$ and width $q$. For each element, the normalized voltage drop is $1/p$ and the corresponding normalized current is $1/q$. 

As shown in figure~\ref{fig:net_pxq}, the rectangular network can be transformed into an equivalent one with fewer elements, by substituting a ``square'' of $k\times k$ elements with a single element. This new element will sustain a normalized current $k/q$ and a voltage drop $k/p$ will develop on it~\cite{Cannon:1992}. The substitution process can be iterated, and any remaining square of elements having side $k>1$ can be substituted with a single element. 

The reader can appreciate that the problem of finding equivalent networks has been related to a geometrical problem: \textbf{given a rectangle having integer sides $p$ and $q$, find a tiling of the rectangle with squares of integer sides}~\cite{Brooks:1940,Bouwkamp:1946,Cannon:1992}. 

\begin{figure}
 	\centering
	\includegraphics{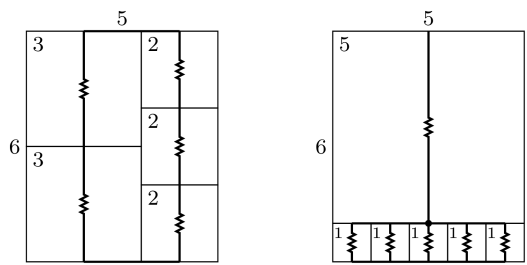}
 	\caption{Tilings of the same $6 \times 5$ rectangle corresponding to the networks of figure~\ref{fig:net_6x5}.}
 	\label{fig:net_6x5_tiled}
\end{figure}

For example, figure~\ref{fig:net_6x5_tiled} shows the two tilings of a $6\times 5$ rectangle corresponding to the two different networks of figure~\ref{fig:net_6x5}; each square of side $k=1,\ldots, 5$ corresponds a single element, which sustains a voltage $k/5$ and a current $k/6$.

The problem of finding a network with value $\rho=p/q$ having a minimal number of elements is therefore related to the problem of finding a tiling of the rectangle with integer sides $p,q$ with the minimum number $n^*(p,q)$ of integer-sided squares. This problem has been widely studied (see e.g.\ \cite{Brooks:1940,Bouwkamp:1946,Cannon:1992,Kenyon:1996,Felgenhauer:2013}) and the following properties of $n^*(p,q)$ are known~\cite{Kenyon:1996,Felgenhauer:2013}:
\begin{enumerate}
\item $n^*(p,q)\geq \max\{p/q,q/p,\log_2 p,\log_2 q\}$;
\item $n^*(p,q)\leq \alpha_0+\alpha_1+\ldots+\alpha_m\leq \max(p,q)$, where $\alpha_0,\ldots,\alpha_m$ are the elements of the continued fraction expansion of $p/q$;
\item\label{property:splitting} $n^*(p+q,q) = n^*(p,q)+1$, if $3p\geq q^2$; or, symmetrically, $n^*(p,p+q) = n^*(p,q)+1$, if $3q\geq p^2$. This property means that for long and thin rectangles for which one of the given conditions is met, the tiling problem can be reduced to that of a smaller rectangle because the tiling includes at least a square with side equal to the shortest side of the original rectangle. 
\end{enumerate}

Even though a general solution to the minimal tiling problem is not yet known, there exists an algorithm that performs an exhaustive search~\cite{Felgenhauer:2013,OEIS:A219158}. To date, all the solutions for $p,q\leq 300$ are known. For larger rectangles, even by taking into account property~(\ref{property:splitting}) to reduce the problem, exhaustive search becomes unfeasible because extremely time-consuming. To overcome this issue, a possibility is that of weakening the condition required in property~(\ref{property:splitting}), so that the original rectangle can be reduced as much as possible before applying exhaustive search, at the cost of accepting possibly non minimal solutions. In this work, where needed, on the basis of~\cite{Kenyon:1996}, the following weakened form of property~(\ref{property:splitting}) is considered\footnote{The fact that property~(\ref{property:splitting_weak}) can lead to non minimal solutions can be demonstrated by considering a $112\times 53$ rectangle for which $n^*(112,53) = 11 < n^*(59,53)+1 = 12$~\cite{Felgenhauer:2013}.}
\begin{enumerate}
\renewcommand{\theenumi}{\roman{enumi}$'$}
\renewcommand{\labelenumi}{(\theenumi)}
\setcounter{enumi}{2}
\item\label{property:splitting_weak} $n^*(p+q,q) = n^*(p,q)+1$, if $p\geq q$; or, symmetrically, $n^*(p,p+q) = n^*(p,q)+1$, if $q\geq p$.
\end{enumerate}
 
Finally, it is worth noting that, for a network with normalized resistance $\rho = p/q$ and a certain associated tiling, the ratio $F$ between the maximum and the minimum currents in the network is given by the ratio of the side of the largest square of the tiling to that of the smallest square. This follows from the fact noted above that the normalized current which flows through an element corresponding to a square of side $k$ is $k/q$. The ratio $F$ is an additional figure-of-merit associated to a network~\cite{Oe:2013}.

\section{Cases}
In QHARS realized with GaAs technology, the plateau index of interest is $i=2$, so $R_\textup{H} = R_\textup{K}/2$. The recommended value of the the von Klitzing constant is $R_\textup{K} = \SI{25812.8074434(84)}{\ohm} \quad [\num{3.2e-10}]$, giving $R_\textup{H} = R_\textup{K}/2 = \SI{12906.4037217(42)}{\ohm}$ \cite{CODATA:2010}. The conventional value adopted internationally for realizing representations of the ohm is $R_{\textup{K-90}} = \SI{25812.807}{\ohm\of{90}}$, giving $R_{\textup{H}-90} = R_{\textup{K-90}}/2 = \SI{12906.4035}{\ohm\of{90}}$. The relative difference between $R_{\textup{H}}$ and $R_{\textup{H-90}}$ is \num{1.718(32)e-8}. In the following, the normalization resistance employed to compute $\rho_0$ and $\delta$ is $R_{\textup{H-90}}$; however, since approximations $\rho$ to $\rho_0$ are considered of interest if $|\delta| <\num{1e-6}$, the results maintain their validity when $R_\textup{H}$ is instead considered.

We investigate here approximations of decadic resistance values, in particular the values \SI{100}{\ohm}, \SI{1}{\kilo\ohm}, \SI{10}{\kilo\ohm}, \SI{100}{\kilo\ohm} and \SI{1}{\mega\ohm}. These values are often those for which a set of artifact resistance standards is maintained, are typically calibrated for dissemination, and are the goal of previous works on QHARS design~\cite{Poirier:2004,Oe:2008,Oe:2010,Oe:2011,Oe:2013}. With available QHARS of these values, the dissemination process could proceed mainly by substitution or $1:1$ comparison calibrations, which do not require the availability of resistance ratio standards that can affect the calibration uncertainty.

Table~\ref{tab:summary} gives a summary of the results of this investigation. For each decadic value $R_0$, the normalized ratio $\rho_0 = R_0/R_\textup{K-90}$ is evaluated. A sequence of rational approximations of $\rho_0$ is generated from the Stern-Brocot tree. Some cases, selected among those with a maximum error of \num{E-6} or better, are developed\footnote{Some of the approximations were previously investigated in~\cite{Oe:2011,Oe:2013}, also. Table~\ref{tab:summary} reports the number of elements in the networks there considered, and the corresponding $F$ ratio.}; one case ($\rho = 203/262$) is taken as example to outline the complete analysis. 

\begin{table}
\caption{Summary of the cases analyzed in this work: $\rho_0 = R_0/R_\textup{H}$ is the normalized target value, $\rho = p/q$ is a rational approximation for $\rho_0$ and $\delta = (\rho-\rho_0)/\rho_0$ is the relative error of the approximation. For each approximation, the last four columns report the properties of the corresponding solution obtained with the algorithm described in section~\ref{sec:qhars_synthesis}: $n$ is the number of elements composing the network and $F$ is the ratio between the maximum and the minimum currents. For some solutions, the last column gives a reference to the corresponding network topology in figure~\ref{fig:net_summary}. Where applicable, for comparison, the results from~\cite{Oe:2013} are also reported.}
\label{tab:summary}
\begin{indented}
\item[]\begin{tabular}{@{}*{8}{l}}
\br
&&\centre{2}{Results from~\cite{Oe:2013}}&\centre{4}{This work}\\ 
\ns 
&&\crule{2}&\crule{4}\\
$\rho$&$\delta$&$n$&$F$&$n$&$F$&Minimal&Figure\\
\mr
\centre{8}{$R_0 = \SI{100}{\ohm}$} \\
\mr
$47/6066$&\num{+1.6e-6}&&&137&47/4&No&\ref{fig:net_summary}\subref{fig:net_47x6066}\\
$78/10067$&\num{-5.2e-7}&151&78&138&78&No&\ref{fig:net_summary}\subref{fig:net_78x10067}\\
$125/16133$&\num{+2.7e-7}&&&139&125/7&No&\\
$203/26200$&\num{-3.4e-8}&&&140&203/10&No&\\
\mr
\centre{8}{$R_0 = \SI{1}{\kilo\ohm}$} \\
\mr
$203/2620$&\num{-3.4e-8}&&&24&203/12&No&\ref{fig:net_summary}\subref{fig:net_203x2620}\\                        
$235/3033$&\num{+1.6e-6}&30&47&24&235/17&No&\ref{fig:net_summary}\subref{fig:net_235x3033}\\
\mr
\centre{8}{$R_0 = \SI{10}{\kilo\ohm}$} \\
\mr
$203/262$&\num{-3.4e-8}&16&67/11&12&119/10&Yes&\ref{fig:net_summary}\subref{fig:net_203x262}\\
\mr
\centre{8}{$R_0 = \SI{100}{\kilo\ohm}$} \\
\mr
$1015/131$&\num{+3.4e-8}&23&131/2&18&131/4&No&\ref{fig:net_summary}\subref{fig:net_1015x131}\\
\mr
\centre{8}{$R_0 = \SI{1}{\mega\ohm}$} \\
\mr
$4029/52$&\num{+1.9e-6}&&&88&52/2&No&\\                                 
$6121/79$&\num{-1.2e-6}&&&87&79/2&No&\ref{fig:net_summary}\subref{fig:net_6121x79}\\                            
$10150/131$&\num{+3.4e-8}&98&131&88&131/2&No&\ref{fig:net_summary}\subref{fig:net_10150x131}\\
\br
\end{tabular}
\end{indented}
\end{table}

\begin{figure}
 	\centering
	\includegraphics{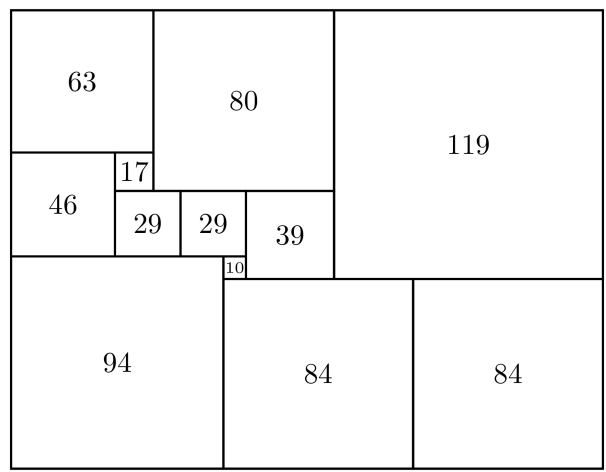}
 	\caption{Minimal tiling of the 262-by-203 rectangle.}
 	\label{fig:rect_203_262_tiled}
\end{figure}

The rectangle corresponding to each approximation $\rho$ is tiled by squares, using the method described in section~\ref{sec:qhars_synthesis}. Figure~\ref{fig:rect_203_262_tiled} shows the tiling found for the example case $203/262$, which corresponds to a minimal tiling (for the other cases, it might occur that further simpler tilings exist). For each case, the number $n$ of QHE network elements involved, the $F$ ratio value and the network topology can be directly deduced from the tiling. 

\begin{figure}[p]
\centering
\subfigure[$47/6066$]{\label{fig:net_47x6066}\includegraphics{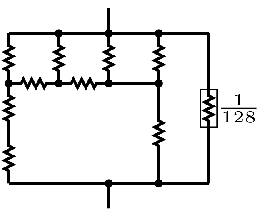}}
\hspace{8ex}
\subfigure[$78/10067$]{\label{fig:net_78x10067}\includegraphics{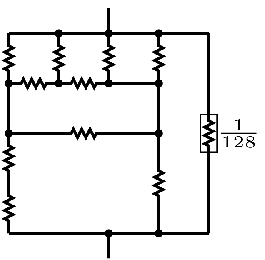}}
\\
\subfigure[$203/2620$]{\label{fig:net_203x2620}\includegraphics{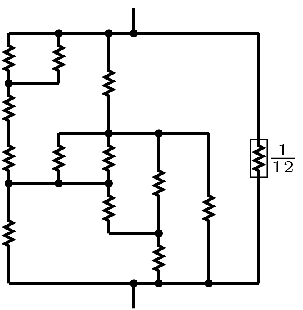}}
\hspace{8ex}
\subfigure[$235/3033$]{\label{fig:net_235x3033}\includegraphics{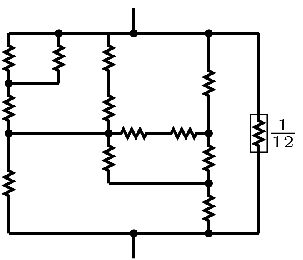}}
\\
\subfigure[$203/262$]{\label{fig:net_203x262}\includegraphics{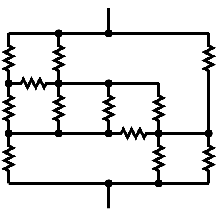}}
\hspace{8ex}
\subfigure[$1015/131$]{\label{fig:net_1015x131}\includegraphics{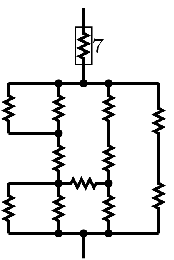}}
\\
\subfigure[$6121/79$]{\label{fig:net_6121x79}\includegraphics{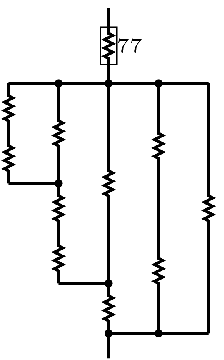}}
\hspace{8ex}
\subfigure[$10150/131$]{\label{fig:net_10150x131}\includegraphics{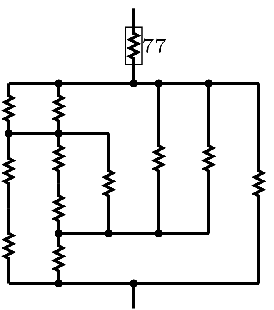}}
\caption{Network topologies corresponding to the cases considered in table~\ref{tab:summary}.}
\label{fig:net_summary}
\end{figure}

Figure~\ref{fig:net_summary} shows the network topologies found for several interesting cases. Each topology is represented as a two-terminal network composed of two-terminal elements. The actual circuit, instead, is provided with four terminals (two current terminals and two voltage ones) and is composed of multiterminal QHE elements. In order to reject the effect of the inevitable contact and wiring resistances, the connection among QHE elements should be realized as multiterminal \emph{multiple-series}, \emph{multiple-parallel}~\cite{Delahaye:1993} or \emph{multiple-bridge}~\cite{Callegaro:2013} connections. 

\begin{figure}
 	\centering
	\includegraphics{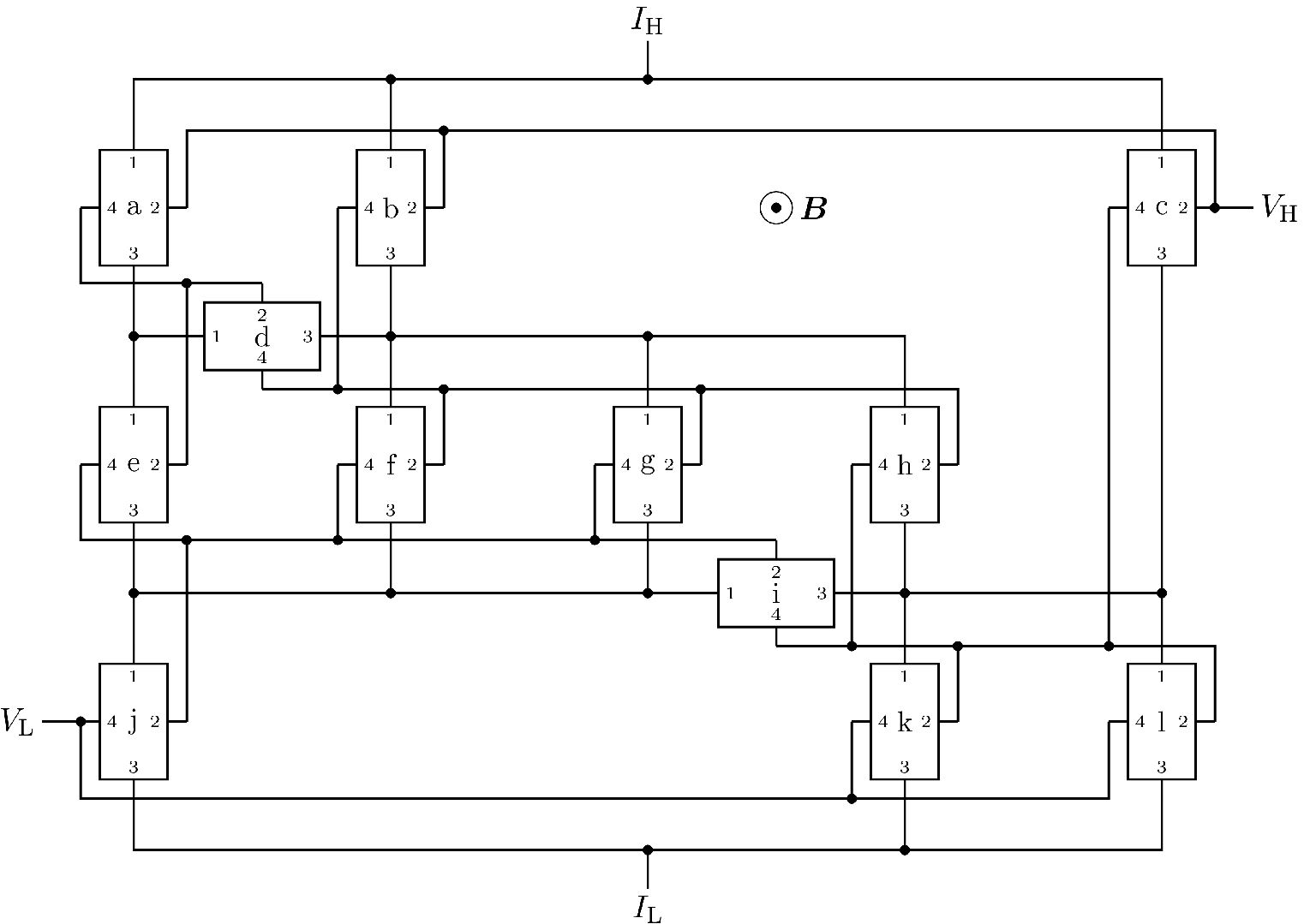}
 	\caption{Circuit diagram corresponding to the tiling of figure~\ref{fig:rect_203_262_tiled} employing 4-terminal QHE elements in multiple-series, multiple-parallel and multiple-bridge connections.}
 	\label{fig:net_203_262_bridge_dc}
\end{figure}

The circuit diagram for the example case $203/262$ is given in figure~\ref{fig:net_203_262_bridge_dc}, where double-series, double-parallel and double-bridge connections are employed. The analysis carried out with the indefinite admittance matrix method of~\cite{Ortolano:2012} yielded a four-terminal resistance $\rho = 203/262 + O(\epsilon^2)$, where $\epsilon$ is the maximum of the set of the normalized parasitic resistance values associated with contacts and wiring (see~\cite{Ortolano:2012,Callegaro:2013} for details).

\section{Conclusions}
We have shown that the synthesis of a QHARS with a given value can be traced back to two mathematical problems, well known and widely developed in literature: i) find a rational approximation of the ratio between the QHARS desired value and the quantized resistance; ii) find an efficient tiling of the rectangle representing the approximation fraction with a small number of squares; the tiling will give the number of QHE elements needed and the way to connect them.

A number of cases, corresponding to decadic QHARS resistance values, have been explicitly considered. The resulting networks have a number of elements significantly less than those given in literature, and should permit an easier realization of integrated QHARS.

The authors are interested in collaborations with QHARS manufacturers to analyze further cases of practical interest.

\section*{References}
\providecommand{\newblock}{}


\begin{thebibliography}{10}
\expandafter\ifx\csname url\endcsname\relax
  \def\url#1{{\tt #1}}\fi
\expandafter\ifx\csname urlprefix\endcsname\relax\def\urlprefix{URL }\fi
\providecommand{\eprint}[2][]{\url{#2}}

\bibitem{Piquemal:1999}
Piquemal F~P~M, Blanchet J, G\`{e}neves G and Andr\'{e} J~P 1999 {\em IEEE
  Trans. Instr. Meas.\/} {\bf 48} 296--300

\bibitem{Poirier:2002}
Poirier W, Bounouh A, Hayashi H, Fhima H, Piquemal F, G\`{e}neves G and
  Andr\'{e} J~P 2002 {\em J. Appl. Phys.\/} {\bf 92} 2844--2854

\bibitem{Bounouh:2003}
Bounouh A, Poirier W, Piquemal F, G\`{e}neves G and Andr\'{e} J~P 2003 {\em
  {IEEE} Trans. Instr. Meas.\/} {\bf 52} 555--558

\bibitem{Poirier:2004}
Poirier W, Bounouh A, Piquemal F and Andr\'{e} J~P 2004 {\em Metrologia\/} {\bf
  41} 285

\bibitem{Hein:2004}
Hein G, Schumacher B and Ahlers F~J 2004 Preparation of quantum {H}all effect
  device arrays {\em 2004 Conference on Precision Electromagnetic Measurements
  Digest\/} pp 273--274

\bibitem{Oe:2008}
Oe T, Kaneko N, Urano C, Itatani T, Ishii H and Kiryu S 2008 Development of
  quantum {H}all array resistance standards at {NMIJ} {\em 2008 Conference on
  Precision Electromagnetic Measurements Digest\/} pp 20--21

\bibitem{Oe:2010}
Oe T, Matsuhiro K, Urano C, Fujino H, Ishii H, Itatani T, Sucheta G, Maezawa M,
  Kiryu S and Kaneko N 2010 Development of 10 k${\Omega}$ quantum {H}all array
  resistance standards at {NMIJ} {\em 2010 Conference on Precision
  Electromagnetic Measurements Digest\/} pp 619--620

\bibitem{Oe:2011}
Oe T, Matsuhiro K, Itatani T, Gorwadkar S, Kiryu S and Kaneko N 2011 {\em IEEE
  Trans. Instr. Meas.\/} {\bf 60} 2590--2595

\bibitem{Konemann:2011}
Konemann J, Ahlers F~J, Pesel E, Pierz K and Schumacher H 2011 {\em IEEE Trans.
  Instr. Meas.\/} {\bf 60} 2512--2516

\bibitem{Woszczyna:2012}
Woszczyna M, Friedemann M, Dziomba T, Weimann T and Ahlers F~J 2011 {\em Appl.
  Phys. Lett\/} {\bf 99} 022112 (pages~3)

\bibitem{Oe:2013}
Oe T, Matsuhiro K, Itatani T, Gorwadkar S, Kiryu S and Kaneko N 2013 {\em IEEE
  Trans. Instr. Meas.\/} {\bf 62} 1755--1759

\bibitem{Ortolano:2012}
Ortolano M and Callegaro L 2012 {\em Metrologia\/} {\bf 49} 1

\bibitem{Kroupa:1974}
Kroupa V~F 1974 {\em IEEE Tran. Instr. Meas.\/} {\bf 23} 521--524

\bibitem{Stern:1858}
Stern M~A 1858 {\em Journal f\"ur die reine und angewandte Mathematik\/} {\bf
  55} 193--220
  \urlprefix\url{http://www.digizeitschriften.de/dms/img/?PPN=GDZPPN002150301}

\bibitem{Brocot:1862}
Brocot A 1862 {\em Calcul des rouages par approximation, nouvelle m\'ethode\/}
  (Paris: Achille Brocot)
  \urlprefix\url{http://gallica.bnf.fr/ark:/12148/bpt6k1661912}

\bibitem{Graham:1994}
Graham R~L, Knuth D~E and Patashnik O 1994 {\em Concrete mathematics\/}
  (Addison-Wesley Professional)

\bibitem{Olds:1963}
Olds C~D 1963 {\em Continued fractions\/} (New York: Random House)

\bibitem{Amengual:2000}
Amengual A 2000 {\em Am. J. Phys.\/} {\bf 68} 175--179

\bibitem{Khan:2012}
Khan S~A 2012 {\em Resonance\/} {\bf 17} 468--475

\bibitem{OEIS:A153588}
{OEIS Foundation Inc} 2013 The on-line encyclopedia of integer sequences
  sequence A153588 \urlprefix\url{http://oeis.org/A153588}

\bibitem{Cannon:1992}
Cannon J~W, Floyd W~J and Parry W~R 1992 Squaring rectangles: the finite
  {R}iemann mapping theorem {\em The Mathematical Legacy of {W}ilhelm {M}agnus:
  Groups, Geometry, and Special Functions : Proceedings of the Conference on
  the Legacy of {W}ilhelm {M}agnus\/} ed Abikoff W, Birman J~S and Kuiken K
  (Brooklyn, NY, US: American Mathematical Society)

\bibitem{Brooks:1940}
Brooks R~L, Smith C~A~B, Stone A~H and Tutte W~T 1940 {\em Duke Math. J\/} {\bf
  7} 312--340

\bibitem{Bouwkamp:1946}
Bouwkamp C~J 1946 On the dissection of rectangles into squares ({I}--{III})
  {\em Koninklinjke Nederlandsche Akademie van Wetenschappen, Proceedings\/}
  vol 49 and 50

\bibitem{Kenyon:1996}
Kenyon R 1996 {\em Journal of Combinatorial Theory, Series A\/} {\bf 76}
  272--291

\bibitem{Felgenhauer:2013}
Felgenhauer B 2013 Filling rectangles with integer-sided squares
  \urlprefix\url{http://int-e.eu/\~{}bf3/squares/}

\bibitem{OEIS:A219158}
{OEIS Foundation Inc} 2013 The on-line encyclopedia of integer sequences
  sequence A219158 \urlprefix\url{http://oeis.org/A219158}

\bibitem{CODATA:2010}
Mohr P~J, Taylor B~N and Newell D~B 2012 {\em Rev. Mod. Phys.\/} {\bf 84}
  1527--1605

\bibitem{Delahaye:1993}
Delahaye F 1993 {\em J. Appl. Phys.\/} {\bf 73} 7914--7920

\bibitem{Callegaro:2013}
Callegaro L and Ortolano M Bridge connection of quantum {H}all elementary
  devices submitted to Metrologia
  \urlprefix\url{http://arxiv.org/abs/1310.5583}

\end{thebibliography}
\end{document}